\documentstyle{article}
\begin{document}
\begin{flushright}
ULB-TH-99-28/\\
hep-th/9911109\\
November 1999
\end{flushright}

\begin{center}
{\Large \bf Interactions of chiral two-forms}\footnote{Talk given
at the TMR-meeting ``Quantum aspects of gauge theories, supersymmetry and unification'',
ENS (Paris), September 1-7, 1999 and at the ``9th Midwest Geometry Conference´´, University of Missouri (Columbia), November 5-7, 1999.
From joint work with M. Henneaux and A. Sevrin.
} \\

\vspace{8mm}

%                      author/address

XAVIER BEKAERT

\end{center}

%                       Abstract

\begin{abstract}
Two issues regarding the interactions of the chiral two-forms
are reviewed.
First, the problem of constructing Lorentz-invariant
self-couplings of a single chiral two-form is investigated in the light of the Dirac-Schwinger
condition on the energy-momentum tensor commutation relations.
We show how the Perry-Schwarz condition follows
from the Dirac-Schwinger criterion and point out that consistency
of the gravitational coupling is automatic.
Secondly, we study the possible local deformations of chiral two-forms.
This problem reduces to the study of the local BRST cohomological group at ghost number zero.
We proof that the only consistent deformations of a system of free chiral two-forms
are (up to redefinitions) deformations that do not
modify the abelian gauge symmetries of the free theory.
The consequence of this result for a system consisting of a number of parallel M5-branes
is explained.
\end{abstract}

%                        Subject

\vspace{5mm}

{\bf Mathematics Subject Classification:} 81T30, 83C99, 81T70

\vspace{3mm}

{\bf Keywords:} Chiral p-forms, Poincar\'e invariance, BRST Cohomology, M5-brane

\vspace{2mm}

%                         Text

\section{Introduction: the chiral two-form and the M5-brane}

In the limit where bulk gravity decouples, the M5-brane is described by a six-dimensional field
theory. Its bosonic sector includes, besides the five scalar fields  which
describe the position of the brane in transverse space, a {\it chiral two-form}
(i.e. a two-form the field strength of which is self dual).

A single M5-brane with strong classical fields is well understood;
its Lagrangian is described in \cite{PST}.
For a long time, the obstacle to get such an action was due to the presence of the chiral two-form.
When one tries to tries to incorporate the self-duality condition into an action a problem arises
with preserving Lorentz invariance. There is a simple non-manifestly Lorentz-invariant free Lagrangian which has been given
in \cite{HT}, and which generalizes the Lagrangian of \cite{FJ} for
chiral bosons (the couplings to gravitation are included).
As a non linear generalization, a Born-Infeld-like action for for a self-interacting chiral two-form
was proposed in \cite{PS}. The couplings to gravitation were added in a second paper \cite{S}.
A manifestly covariant formulation of chiral $p$-forms has been developed in an interesting series of papers \cite{PST2}.
This formulation is characterized by the presence of an extra field and an extra gauge
invariance. This extra field occurs non-polynomially in the action, even for the
free chiral $p$-forms. It has been shown to be equivalent
to the non-manifestly covariant treatment of \cite{HT} in Minkowski space \cite{PST2}.

The question of Lorentz-invariant
self-couplings (as well as consistent self-couplings in an external
gravitational background) for a single chiral $2$-form is the subject of the section \ref{DiracSchw}.
We show that this question can be handled by means of the Dirac-Schwinger condition on the commutation
relations of the components of the energy-momentum tensor \cite{DirSchw}.
This condition leads directly to the differential
equation obtained in \cite{PS} and implies automatically consistency of
the gravitational coupling \cite{bekaert}.

When several, say $n$, M5-branes coincide,
little is known\footnote{There are several indications that this is a highly unusual  system.
Both entropy considerations \cite{kleb} and the calculation of the
conformal  anomaly of the partition function \cite{kostas} show that the
theory should have $n^3$ instead of  $n^2$ degrees of freedom.}. Compactifying one dimension of M-theory on  a circle yields
type IIA string theory.
If the  branes are longitudinal to the compact
direction, the M5-branes appear as a set  of coinciding D4-branes which are
quite well understood. Their dynamics is  governed by a five-dimensional
$U(n)$ Born-Infeld theory which, ignoring  higher derivative terms, is an ordinary
$U(n)$ non-abelian gauge theory.  Turning back to the eleven dimensional
picture, this suggests that a non-abelian  extension of the chiral two-form
should exist.

Deformations of a system of $N$ free chiral two-forms in six dimensions are only considered
($N$ is expected to grow like $n^3$).
We ignore the fermions and the scalar fields as we believe that they will not modify our conclusions\footnote{
In fact this can easily be proven for the scalar fields because they are inert under the two-form gauge symmetry.}.
The analysis, outlined in the section \ref{deformations}, shows
that all continuous and local deformations of the free chiral 2-forms
does not modify the gauge algebra which remains abelian \cite{BHS}.
As a consequence, {\it no local field theory
continuously connected with the free theory can describe a system of $n$ coinciding $M5$-branes}.
This does not exclude non-local deformations of the abelian theory or
local lagrangians that cannot be continuously deformed to the free one.

\section{Covariant self interactions of a single chiral two-form}\label{DiracSchw}

The non manifestly covariant action for a chiral 2-form in a gravitational background is
\begin{equation}
S[A_{ij}] = \int dx^0 d^5x B^{ij} \partial_0 A_{ij} - \int dx^0 H \; \; \;
(i,j, \dots = 1, \dots, 5)  \label{action0}
\end{equation}
with
\begin{equation}
B^{ij} = \frac{1}{3!} \epsilon^{ijklm} F_{klm}, \; \; F_{\mu \nu \lambda} =
\partial_\mu A_{\nu \lambda} - \partial_{\nu} A_{\mu \lambda} -
\partial_\lambda A_{\nu \mu}  \label{defB}
\end{equation}
and
\begin{equation}
H = \int d^5x ( N {\cal H} + N^k {\cal H}_k).
\end{equation}
Here, $N$ and $N^k$ are the standard lapse and shift. The
magnetic field $B^{ij}$ is a spatial tensor density of weight one.
In the absence of self-interactions, the energy density $%
{\cal H}$ is given by \cite{HT}
\begin{equation}
{\cal H} = \frac{1}{\sqrt{g}} B^{ij} B_{ij}  \label{Hbis}
\end{equation}
where the spatial indices are lowered and raised with the spatial metric and
its five-dimensional inverse, while $g$ is the determinant of $g_{ij}$. The
energy density generates displacements normal to the slices of constant $x^0$.
The momentum density ${\cal H}_k$, on the other hand, is purely
kinematical and generates tangent displacements. It is explicitly given by
\begin{equation}
{\cal H}_k = \frac{1}{2} \epsilon_{ijmnk}B^{ij}B^{mn}.
\end{equation}

In order to write the action (\ref{action0}), it is necessary to assume that
spacetime has the product form $T \times \Sigma$ where $T$ is the manifold of the time variable (usually a line).
Of course, a spatial coordinate could equivalently play the r\^ole of the time variable, as in \cite{PS}.

Since $B^{ij}$ is gauge-invariant and identically transverse ($\partial _i
B^{ij}=0$), the action (\ref{action0}) is manifestly invariant under the usual gauge
transformations $\delta_{\Lambda} A_{ij} = \partial_i \Lambda_j - \partial_j \Lambda_i$.
In flat space, it is also invariant under Lorentz transformations, but
these do not take the usual form \cite{HT}.

We define the exterior form $B$ to be the (time-dependent) spatial $2$-form
with components $B_{ij}/\sqrt{g}$. The equations of motion that follow from
the action are $\tilde{d}[N(E-B)] = 0$, where $E$ is the electric spatial $2$-form defined through
\begin{equation}
E_{ij} \equiv \frac{\dot{A}_{ij} - N^s F_{sij}}{N}  \label{defE}
\end{equation}
and where $\tilde{d}$ is the spatial exterior derivative operator.
In the case where the second Betti number $b_2(\Sigma)$ of the spatial sections vanishes, this
equation implies $N(E-B)=\tilde{d}m$, where $m$ is an arbitrary spatial $1$-form. To
bring this equation to a more familiar form, one sets $m_i = A_{0i}$. The
equations of motion read then $F= \! ^{*} F$,
where $F_{0ij} = \dot{A}_{ij} - \partial _i A_{0j} + \partial _j A_{0i}$.
This is the standard self-duality condition. Alternatively, one may use the
gauge freedom to set $m=0$, which yields the self-duality condition in the
temporal gauge ($E=B$).

The brackets of the gauge-invariant magnetic fields $B^{ij}$
can be found by follow the Dirac method for constrained systems \cite{DMethod} but one may shortcut
the whole procedure and directly read the brackets from the kinetic term of the action (\ref
{action0})
\begin{equation}
[B^{ij}({\bf x}), B^{mn}({\bf x^{\prime}})] = \frac{1}{4} \epsilon^{ijmnk}
\delta,_k({\bf x} - {\bf x^{\prime}}).  \label{brackets}
\end{equation}

A direct calculation of the brackets of the energy densities ${\cal H}({\bf x})$
at two different space points using only the form of $%
{\cal H}$ and the brackets (\ref{brackets}) yields
\begin{equation}
[{\cal H}({\bf x}), {\cal H}({\bf x^{\prime}})] = ({\cal H}^k({\bf x}) +
{\cal H}^k({\bf x^{\prime}})) \delta,_k({\bf x} - {\bf x^{\prime}}).
\label{DS}
\end{equation}

Now, we can ask ourselves how can we introduce interactions preserving Lorentz invariance ?
When one can use the tensor calculus, it is rather easy to construct
interactions that preserve Lorentz invariance\footnote{These interactions should
also preserve the number of (possibly deformed) gauge symmetries (if any),
but this aspect is rather immediate for $p$-form gauge symmetries -
although it is less obvious for the extra gauge symmetry of \cite{PST2}.}.
But there is an alternative way to control Lorentz invariance. It is through the
commutation relations of the energy-momentum tensor components. Because the
energy-momentum tensor is the source of the gravitational field, the method
gives at little extra price a complete grasp on the gravitational
interactions. As shown by Dirac and Schwinger \cite{DirSchw}, a
sufficient condition for a manifestly rotation and translation invariant
theory (in space) to be also Lorentz-invariant is that its energy density
fulfills the commutation relations (\ref{DS}). The condition is necessary
when one turns to gravitation \cite{T}. The method is more cumbersome than the tensor
calculus when one can use the tensor calculus, but has the advantage of
being still available even when manifestly invariant methods do not exist.

In the Dirac-Schwinger approach,
the question is to find the most general ${\cal H}$ fulfilling
(\ref{DS}). The energy-density ${\cal H}$ must
be a spatial scalar density
in order to fulfill the kinematical commutation relations
$[{\cal H}({\bf x}),{\cal H}_k({\bf x'})] \sim
{\cal H}({\bf x'})\delta,_k({\bf x} - {\bf x'})$ and depends
on $A_{ij}$ through $B_{ij}$ in order to be
gauge-invariant.
In five dimensions, there are only two independent
invariants that can be
made out of $B_{ij}$, e.g.
\begin{equation}
y_1 = - \frac{1}{2g} B_{ij} B ^{ij}, \; \;
y_2 = \frac{1}{4g^2} B_{ij} B^{jk} B_{km} B^{mi}.
\end{equation}
Set
\begin{equation}
{\cal H}= f(y_1,y_2)\sqrt{g}, \; f_1 = \partial_1 f, \;
f_2 = \partial_2 f.
\end{equation}

Then, the condition (\ref{DS}) yields to
\begin{equation}
f_1^2 + y_1 f_1 f_2 + (\frac{1}{2} y^2_1 - y_2) f_2^2 = 4
\end{equation}
which is precisely the equation found in \cite{PS}. The Dirac-Schwinger criterion yields thus directly the
Perry-Schwarz equation, whose solutions are investigated in \cite{PS,HKS}.

In the flat space context ($g_{ij}= \delta_{ij}$, $N=1$, $N^k=0$),
the equation (\ref{DS}) guarantees that
the interactions are Lorentz-invariant and no further work is
required. It also guarantees complete
consistency in a gravitational background because of
locality of ${\cal H}$ in the metric $g_{ij}$.

\section{Consistent deformations of free chiral two-forms}\label{deformations}

The starting point is the action of a collection of
$N$ free chiral $2$-forms $A^A_{ij}$
($A = 1, \dots, N;i, j= 1, \dots, 5$) in Minkowski spacetime
\begin{equation}
S[A^A_{ij}]= \sum_A\int d^6x (B^{Aij} \dot{A}^A_{ij} - B^{Aij}B^A_{ij})
\label{freeaction}
\end{equation}
which is the sum of $N$ free action of the form (\ref{action0}) in
flat background. The magnetic fields $B^{Aij}$ in (\ref{freeaction}) are defined through
\begin{equation}
B^A_{ij} = \frac{1}{3!} \epsilon_{ijklm} F^{Aklm}, \;
F^A_{ijk} = \partial_i A^A_{jk} + \partial_j A^A_{ki}
+ \partial_k A^A_{ij}.
\end{equation}
The action $S_0$ is invariant under the following abelian gauge transformations
\begin{equation}
\delta_{\Lambda}A^A_{ij}=\partial_i \Lambda^A_j - \partial_j \Lambda^A_i.
\label{cling}\end{equation}
This set of gauge transformations is reducible because if
$\Lambda^A_i = \partial_i \epsilon^A$, the variation of $A^A_{ij}$ is 0.

Our strategy for studying the possible local deformations of the
action (\ref{freeaction}) is based on the observation that these
are in bijective correpondence with the local BRST
cohomological group $H^{0,6}(s \vert d)$ \cite{BH}, where $s$ is the
BRST differential acting on the fields, the ghosts, and their
conjugate antifields, $d$ is the ordinary space-time exterior derivative and
the upper indices refer to ghost number and form degree resp.
By applying the standard method of the antifield formalism \cite{BV},
one can introduce the ghosts $C^A_{i}$, the ghosts of ghosts $\eta^A$,
and the antifields $A^{*Aij}$, $C^{*Ai}$ and $\eta^{*A}$.
Their respective parity, ghost number, antighost number are listed in following table
\begin{table}[h]
\centerline{
\begin{tabular}{|c|c|c|c|c|}
\hline
 & parity & antigh & puregh & gh\\
\hline
$A^A_{ij}$ & 0 & 0 & 0 & 0\\ \hline
$C^A_i$ & 1 & 0 & 1 & 1\\ \hline
$\eta^A$ & 0 & 0 & 2 & 2 \\ \hline
$A^{*Aij}$ & 1 & 1 & 0 & -1 \\ \hline
$C^{*Ai}$ & 0 & 2 & 0 & -2 \\ \hline
$\eta^{*A}$ & 1 & 3 & 0 & -3 \\ \hline
$s$ & 1 & - & - & 1 \\ \hline
$\delta$ & 1 & -1 & 0 & 1 \\ \hline
$\gamma$ & 1 & 0 & 1 & 1 \\ \hline
\end{tabular}
}
\caption{table of the respective parity, ghost number, antighost number of the variables and the differential operators}
\label{variables}
\end{table}

The BRST operator $s$ is given by $s = \delta + \gamma$
with
\begin{eqnarray}
\delta A^A_{ij}&=&\delta C^A_{i}=\delta \eta^A=0,\\
\delta A^{*Aij}&=&2\partial_k F^{Akij}-\epsilon^{ijklm}
\partial_k \dot{A}^A_{lm},\\
\delta C^{*Ai}&=&\partial_j A^{*Aij},\; \;\delta \eta^{*A} = \partial_i C^{*Ai}
\end{eqnarray}
and
\begin{eqnarray}
\gamma A^A_{ij}&=&\partial_i C^A_j - \partial_j C^A_i,\\
\gamma C^A_i&=&\partial_i \eta^A,\; \; \gamma \eta^A = 0, \label{yep}\\
\gamma A^{*Aij}&=&\gamma C^{*Ai}=\gamma \eta^{*A}=0.
\end{eqnarray}
One verifies that $\delta^2=\gamma^2=\delta\gamma+\gamma\delta=0$.

The cocycle condition defining elements of $H^{0,6}(s \vert d)$ is the
``Wess-Zumino condition" at ghost number zero,
\begin{equation}
sa + db = 0, \; \; \; gh(a) = 0.
\label{WZ}
\end{equation}
Any solution of (\ref{WZ}) defines a consistent deformation
of the action  (\ref{freeaction}) through
$S[A^A_{ij}] \rightarrow S[A^A_{ij}] + g \int d^6x a_0$,
where $a_0$ is the antifield-independent component of $a$.
The deformation is consistent to first-order in $g$,
in the sense that one can simultaneously deform the
original gauge symmetry (\ref{cling}) in such a way that the deformed action is invariant under the
deformed gauge symmetry up to terms of order $g$ (included).  The
antifield-dependent components of $a$ contain informations about
the deformation of the gauge symmetry.
Trivial solutions of (\ref{WZ}) are of the form $a = \gamma c + de$ and
correspond to $a_0$'s that can be redefined away through field
redefinitions. Of course, there are also consistency conditions
on the deformations arising from higher-order terms ($g^2$ and
higher), but it turns out that in the case at hand, consistency
to first order already restricts dramatically the possibilities.

In general, there are three possible types of consistent deformations of the action.
First, one may deform the action without
modifying the gauge symmetry.  In that case, $a$ does not
depend on the antifields, $a = a_0$.  These deformations
contain only strictly gauge-invariant terms, i.e., polynomials in
the abelian curvatures and their derivatives (Born-Infeld
terms are in this category) as well as Chern-Simons terms, which are
(off-shell) gauge-invariant under the abelian gauge
symmetry up to a total derivative. An example of a Chern-Simons term
is given by the kinetic term of (\ref{freeaction}), which can be rewritten
as $F \wedge \partial_0 A$.
\footnote{The spatial $2$-forms $A_{ij}^A$ and their successive time derivatives are effectively independent.}
Second, one may deform the action and the gauge transformations while
keeping their algebra invariant.  In BRST terms, the corresponding cocycles
involve (non trivially) the antifields $A^{*Aij}$ but not $C^{*Ai}$ or
$\eta^{*A}$.
Finally, one may deform everything, including the gauge algebra; the
corresponding cocycles involve all the antifields.

Reformulating the problem of deforming the free action (\ref{freeaction})
in terms of BRST cohomology enables one to use the powerful tools  of homological
algebra. Following the approach of \cite{BBH}, we have
completely worked out the BRST cohomogical classes at ghost number zero.
In particular, we have established that one can always get rid of the
antifields by adding trivial solutions \cite{BHS} (The complete proof will be given in \cite{BHS2}).
This result can be straightforwardly generalized for a system of chiral $p$-forms in $2p+2$ dimensions\footnote{
For example, the same techniques have been used in 10 dimensions
to show that the only symmetry-deforming consistent vertex for a
system of one chiral $4$-form and two $2$-forms is the one that occurs 
in the type $II_B$ supergravity Lagrangian \cite{BHS3}. These interactions
deform the gauge transformations of the exterior forms but not their
algebra, which remain abelian.}.
In other words, {\it the only consistent local deformations of a system of free chiral $p$-forms
are (up to redefinitions) deformations that do not
modify the abelian gauge symmetries of the free theory.}
These involve the abelian curvatures or Chern-Simons terms.
There are no other consistent, local, deformations (This completely justify the assumptions made for the dependence
of ${\cal H}$ on the $A_{ij}$ through the $B_{ij}$ in the section \ref{DiracSchw}).

\section{Acknowledgments}

I am grateful to the respective organizers of the meeting ``Quantum aspects of gauge theories, supersymmetry and unification''
and the ``9th Midwest Geometry Conference'' for their very enjoyable conferences.
I would like to thank my collaborators on the works which were the sources
of this talk, M. Henneaux and A. Sevrin. This work has been supported in part by the ``Actions de
Recherche Concert{\'e}es" of the ``Direction de la Recherche
Scientifique - Communaut{\'e} Fran{\c c}aise de Belgique" and by
IISN - Belgium. The author is ``Chercheur I.I.S.N.'' (Belgium).

%                                                       Bibliography

\bibliographystyle{BKstyle}

\vspace{.1cm}

{\bf Adress:}\\
{\em Physique Th\'eorique et Math\'ematique,}\\
{\em Universit\'e Libre de Bruxelles,}\\
{\em Campus Plaine C.P. 231}\\
{\em B-1050 Bruxelles, Belgium} \\
{\em xbekaert@ulb.ac.be}

\end{document}